\documentclass[sigconf]{acmart}
\usepackage{graphicx}
\usepackage{multirow}
\usepackage{booktabs} 
\usepackage{bm}
\usepackage{balance}

\hypersetup{colorlinks,urlcolor=}
\copyrightyear{2017} 
\acmYear{2017} 
\setcopyright{acmcopyright}
\acmConference{MM '17}{October 23--27, 2017}{Mountain View, CA, USA}\acmPrice{15.00}\acmDOI{10.1145/3123266.3123312}
\acmISBN{978-1-4503-4906-2/17/10}

\begin{document}
\title{Region-Based Image Retrieval Revisited}

\author{Ryota Hinami}
\affiliation{\institution{The University of Tokyo}}
\email{hinami@nii.ac.jp}

\author{Yusuke Matsui}
\affiliation{\institution{National Institute of Infomatics}}
\email{matsui@nii.ac.jp}

\author{Shin'ichi Satoh}
\affiliation{\institution{National Institute of Infomatics}}
\email{satoh@nii.ac.jp}


\begin{abstract} 
Region-based image retrieval (RBIR) technique is revisited. In early attempts at RBIR in the late 90s, researchers found many ways to specify region-based queries and spatial relationships; however, the way to characterize the regions, such as by using color histograms, were very poor at that time. 
Here, we revisit RBIR by incorporating semantic specification of objects and intuitive specification of spatial relationships. 
Our contributions are the following. First, to support multiple aspects of semantic object specification (category, instance, and attribute), we propose a multitask CNN feature that allows us to use deep learning technique and to jointly handle multi-aspect object specification. Second, to help users specify spatial relationships among objects in an intuitive way, we propose recommendation techniques of spatial relationships. In particular, by mining the search results, a system can recommend feasible spatial relationships among the objects. The system also can recommend likely spatial relationships by assigned object category names based on language prior. Moreover, object-level inverted indexing supports very fast shortlist generation, and re-ranking based on spatial constraints provides users with instant RBIR experiences.
\end{abstract} 


\maketitle

\section{Introduction} 
Searching images by describing their content, a task known as content-based image retrieval (CBIR), 
is one of the most exciting and successful applications of multimedia processing.
The pipeline of typical CBIR is as follows.
An image in a collection is indexed by its category tag and/or image descriptor.
For example, given an image of a dog, a label `dog' can be assigned,
and/or an image feature vector can be extracted.
In the query phase, the user can search for images by specifying a tag or a query image.
A tag-based search can be performed by simply looking for images with the specified keyword, 
while an image-based search can be accomplished by performing a nearest-neighbor search on feature vectors.

Despite the success of current CBIR systems,
there are three fundamental problems that narrow the scope of image searches.
\textbf{(1) Handling multiple objects:}
Typically, an image is indexed by a global feature that represents a whole image.
This makes it hard to search by making multiple queries with a relationship such as
``a human is next to a dog'' because the global feature does not 
contain spatial information.
\textbf{(2) Specifying spatial relationship:}
Even if multiple objects can be handled by some other means, 
specifying spatial relationship of objects is not straightforward.
Several studies have tried to tackle this problem by using graph-based queries~\cite{Johnson2015a,Prabhu2015}
that represent relationships of objects;
however, such queries are not suitable for end-user applications
because their specification and refinement are difficult.
Several interactive systems~\cite{Smith1996,Chang1997} can specify a simple spatial query intuitively, 
but cannot specify complex relationships between objects.
\textbf{(3) Searching visual concepts:}
Tag-based searches with keyword queries are a simple way to search for images with specific visual concepts,
such as object category (`dog') and attribute (`running'). 
However, managing tags is not so easy if we have to consider multiple objects.
For example, consider an image showing a human standing and a dog running;
it is not clear whether we should assign a tag `running' to this image.
Moreover, the query of a tag-based search is limited to be within the closed vocabulary of the assigned tags,
and annotating images involves huge amounts of manual labor.

\begin{figure}[t] 
\begin{center}
\includegraphics[width=1.00\linewidth]{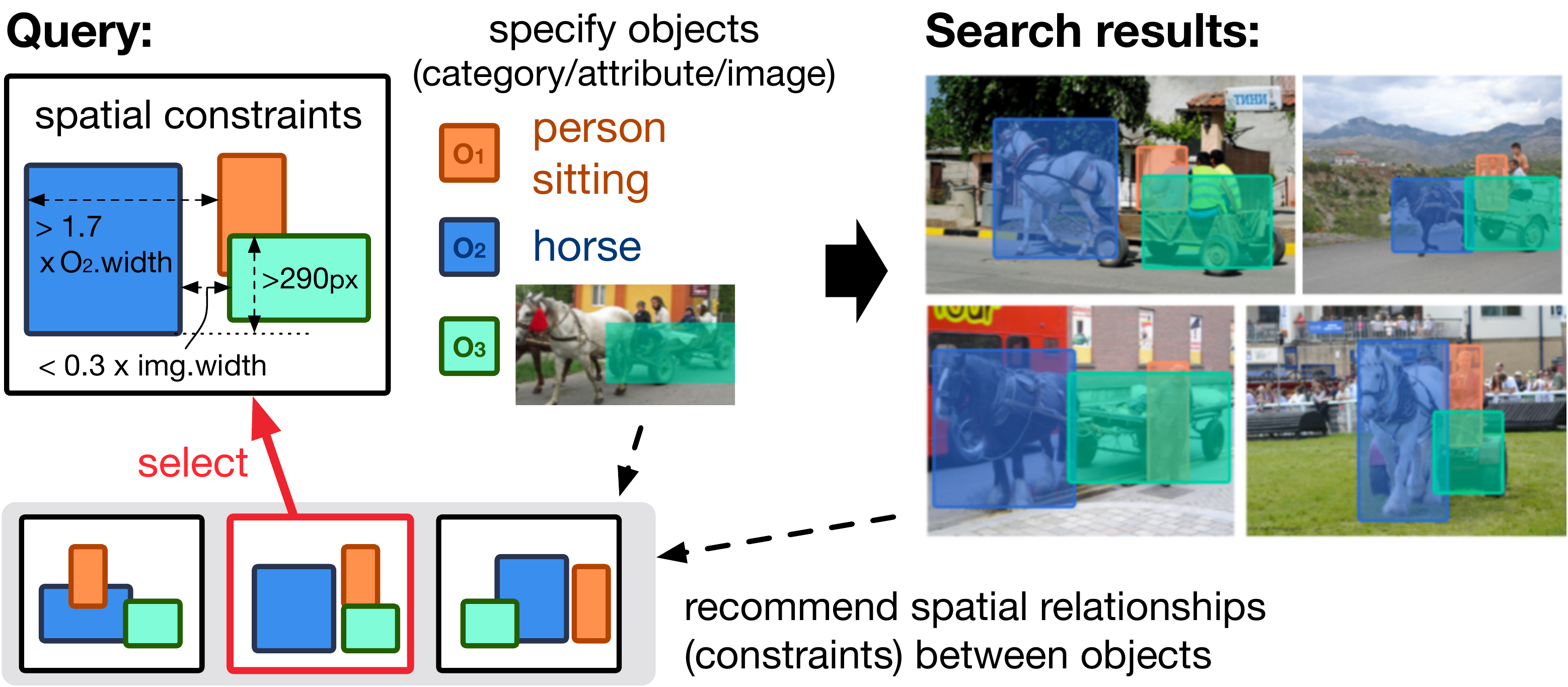}
\vspace{-7mm}
\caption{
Our system enables users to intuitively search the images 
that accord with their complex search intention.
}
\vspace{-6mm}
\label{fig:top} 
\end{center} 
\end{figure}

We created an interactive region-based image retrieval system (Fig.~\ref{fig:top})
that solves the above three problems, as follows:
\begin{enumerate}
\setlength{\leftskip}{-14pt}
	\item 
    Our system provides an interactive spatial querying interface by which
	users can easily locate multiple objects (see the supplemental video).
	Our system is inspired by classical region-based image retrieval (RBIR) systems~\cite{Carson1999}, 
    but is substantially faster, more scalable, and more accurate
    as it takes advantage of recent advances in CNNs and state-of-the-art indexing. 
	\item Our system provides recommendation functions
    that make it easy for users to specify spatial relationships.
	Given initial search results or a query, the system suggests possible 
    spatial relationships among objects.
    Since the suggestions are given as position constraint equations that are unambiguous and understandable, 
	users can interactively refine the suggested query.
	\item Our system incorporates a multitask CNN feature for searching visual concepts,
	which is effective at looking for similar objects and
	discriminative to the object category and attribute.
    By learning classifiers on this feature space,
	users can search for objects by category (`dog') and by attribute (`running') 
    with high accuracy without relying on tag-based searches.
\end{enumerate}

The combination above enables users to search images that match 
their complex search intention.
Figure~\ref{fig:top} shows an example where search intention is
``{\it a person is sitting on horse carriage}''.
The diverse intention can be represented by combining the query of 
category, attribute, and image.
The spatial constraints can be easily specified by interactive 
interface and recommendations.

\section{Related Work}  
{\bf Region-based image retrieval.}
RBIR was proposed as an extension of global feature-based CBIR that 
had received much attention in the late 90s and early 00s~\cite{Carson1999,Wang2001}.
Unlike global feature-based approaches,
RBIR extracts features from multiple sub-regions,
which correspond to objects in the image.
RBIR provided the way to focus on the object details, 
which have much benefit such as that 1) 
the accuracy of queried object was improved by using region-based matching, 
and 2) spatial localization of queried object is easy.


Recently, region-based approach has attracted a lot of attention because of the
success in object detection of R-CNN~\cite{Girshick2014},
which is based on the region proposals~\cite{Uijlings2013} 
and CNN features extracted from the regions.
Several studies~\cite{Bhattacharjee,Xie2015,Bhattacharjee2016,Kiapour2016,Sun2015,Hinami2016} 
incorporated this approach into image retrieval.
They incorporated region-based CNNs that localize objects~\cite{Bhattacharjee,Bhattacharjee2016}
and/or enhance the accuracy of matching images by capturing detailed information on objects ~\cite{Sun2015}.
Hinami and Satoh~\cite{Hinami2016} succeeded in retrieving 
and localizing objects of a certain category (e.g., dog) by indexing region-based CNN features.
The reason for success of region-based CNN features is that they can describe more detailed properties 
of objects than global features can 
and are able to capture higher-level object semantics than local descriptors (e.g., SIFT) can.

{\bf Spatial query and relationship.}
Several early CBIR studies~\cite{Smith1996,Chang1997} investigated interactive specifications of 
spatial queries by using sketches, etc.
VisualSEEk~\cite{Smith1996} searched for images by using spatial queries such as 
those indicating the absolute and relative locations of objects, 
which were specified by diagramming spatial arrangements of regions.
VideoQ~\cite{Chang1997} used sketch-based interactive queries to specify spatiotemporal constraints
on objects, such as motion in content-based video retrieval.
More recent studies~\cite{liu2010robust,xu2010image,longspatial} can handle 
the query composed of semantic concepts (i.e., object categories) and their spatial layout. 
Other studies~\cite{Lokoc2014,Bla2015} demonstrated the effectiveness of 
spatial position-based interactive filtering for video search in the competitions~\cite{Snoek2014}.
Although interactive systems make it easy for the user to specify and refine spatial queries,
they have trouble specifying complex relationships among objects. 

Other approaches~\cite{Johnson2015a,Prabhu2015,Cao2015,Malisiewicz2009,Lan2012,Feris2011} 
encode relationships between objects within a graph or a descriptor,
where images with similar contexts are retrieved by matching them.
The query is represented by graph in \cite{Johnson2015a,Prabhu2015} 
that includes object class information as well as spatial contexts among objects.
Guerrero et al.~\cite{Guerrero2015a} encoded the spatial relationship into the descriptor;
their work shows the effectiveness in describing complex spatial relationship between two objects.
Cao et al.~\cite{Cao2015} encode high-order contextual information among objects into the descriptor 
to measure the strength of interaction, 
where the objects with similar context can be retrieved by the value.
While these approaches can handle complex relationships,
query specification and refinement are generally difficult for the user and not suitable 
for interactive end-user applications.

\begin{figure*}[t] 
\begin{center}
\includegraphics[width=1.00\linewidth]{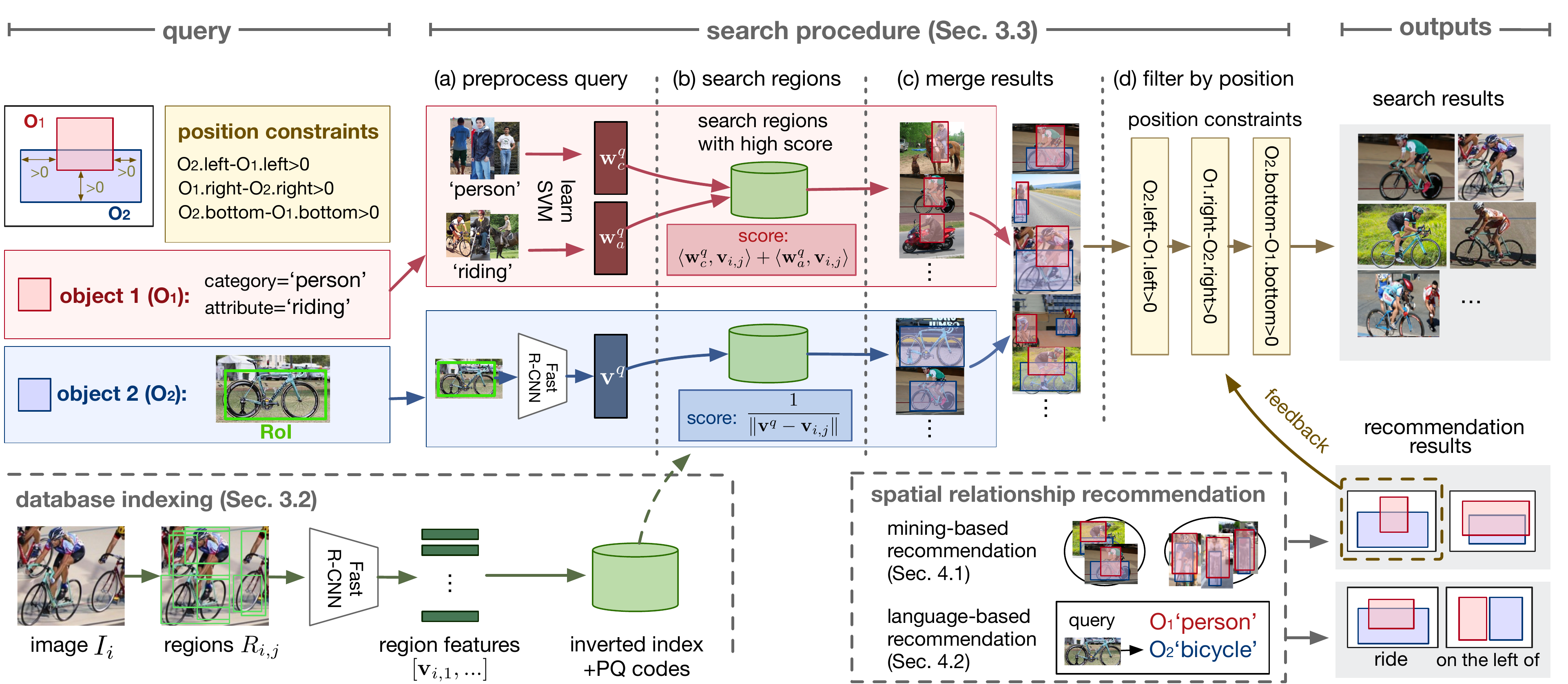}
\vspace{-5mm}
\caption{Pipeline of our region-based image retrieval system.}
\vspace{-2mm}
\label{fig:pipeline} 
\end{center} 
\end{figure*}
\begin{figure}[t] 
\begin{center}
\includegraphics[width=1.00\linewidth]{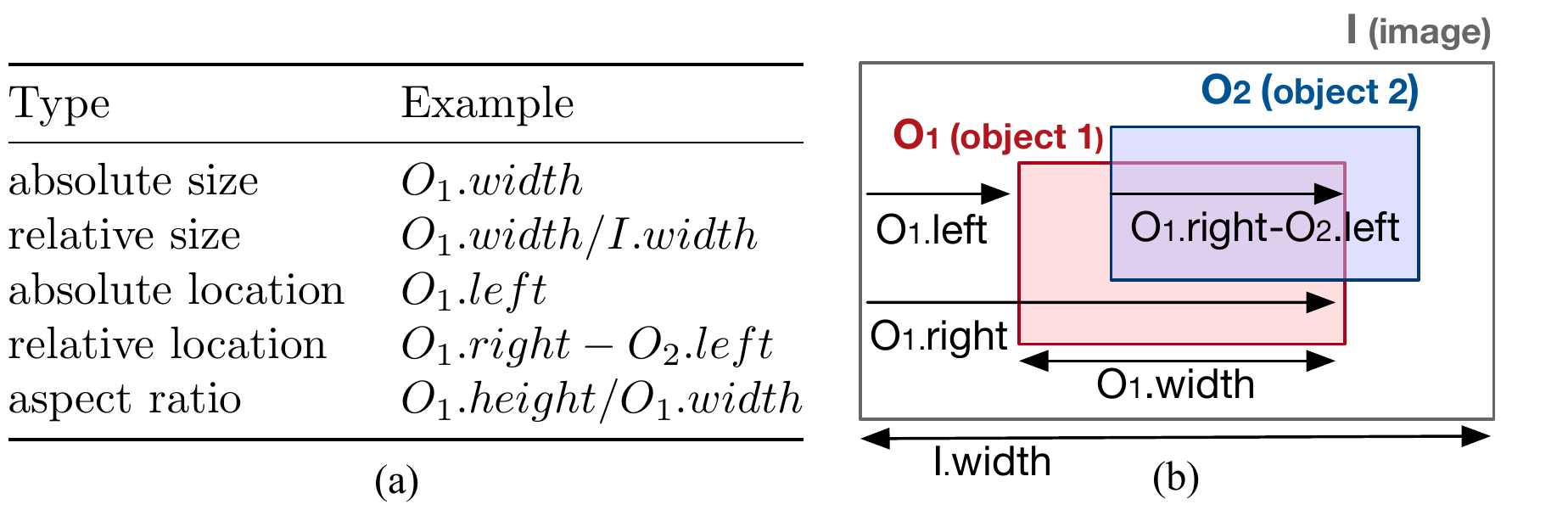}
\vspace{-5mm}
\caption{Position feature (a) examples and (b) illustrations.}
\vspace{-3mm}
\label{fig:posfeat} 
\end{center} 
\end{figure}
\section{The Proposed System: Overview} 
\label{sec:method}
The pipeline of our system is illustrated in Fig.~\ref{fig:pipeline}.
Our system is distinct from typical global feature-based image retrieval;
features of multiple regions in image are indexed following previous RBIR systems, 
where regions generally correspond to the object candidates in the image.
This region-level indexing allows us to 
1) describe the semantics of objects in more detail 
and 2) identify the spatial position of each retrieved object.
The user specifies objects as a query;
our system can take an example image as in previous RBIR systems~\cite{Carson1999},
as well as an object category (e.g., `dog', `person') 
and object attributes (e.g., `running', `white').
The user can retrieve objects in any type of query 
by describing regions with multitask CNN features (see Sec.~\ref{sec:mt}).
Moreover, our system can easily handle queries of multiple objects, 
wherein different types of query can be specified for each object.
In the example shown in Fig.~\ref{fig:pipeline}, 
the number of queried objects $n_o$ is two; 
a category (`person') and an attribute (`riding') are specified in the first object-level query $O_1$,
and an image of a bicycle is specified within an region-of-interest (RoI) in the second object-level query $O_2$.

After the objects are specified, the system retrieves candidate images
and allows users to specify position constraints among the objects.
Our position constraints are specified using a {\it position feature}
that is computed from the geometric locations of the objects 
$O_1$,...,$O_{n_o}$ in the image $I$.
As illustrated in Fig.~\ref{fig:posfeat},
each position feature corresponds to meaningful positional information such as 
object size (e.g., $O_1.width$), distance between objects (e.g., $O_1.left-O_2.right$), 
and aspect ratio (e.g., $O_1.height/O_1.width$).
The number of position feature types $n_p$ is 19, 82, and 213 when $n_o$=1, 2, and 3, respectively
(breakdown of $n_p$=82 when $n_o$=2: size=16, location=32, 
relation between multiple objects=32, and aspect ratio=2).
A position constraint consists of the type of position feature and its threshold 
(the position feature is above or below the threshold).
The spatial location, size, and relationship among objects can be specified 
by appropriately specifying the position constraints;
for example, `$O_1.right/I.width>0.3$' (location in the image), 
`$O_1.width>500px$' (object size), or
`$O_1.left-O_2.right>0$' (relationship between objects).

\subsection{Position Query Specification}
\label{sec:pos}
Our system provides three ways to specify the position constraints.
The first way is to specify the constraints manually 
by the user selecting a feature type and threshold.
This method works well for specifying simple constraints 
such as `$O_1.width/I.width > 0.5$', (e.g., $O_1$ is big), 
or `$O_1.right - O_2.left > 0$', (e.g., $O_1$ is on the left of $O_2$).
However, manual specification is frustrating especially when there are many constraints.

The second way is by using our interactive interface.
The user interactively specifies the position of objects by dragging and resizing the boxes on the canvas,
and the search results are updated immediately in synchronization with the interface.
Our system maps the box positions on the interface into constraints regarding
the four position features that determine the relative position of the region $O_i$ in the image $I$, namely, 
$O_i.left/I.width$, $O_i.right/I.width$, 
$O_i.top/I.height$, and $O_i.bottom/I.height$.
The boxes on the canvas (specified by the user) also represents 
specific values of these four position features.  
To convert these values into position constraints,
each of four position features $x$ is constrained to be 
$0.9x^q$<$x$<$1.1x^q$ (i.e., $x \sim x^q$),
where $x^q$ is the position feature computed from the boxes on the canvas.
Although this specification with an interactive interface is easy and intuitive, 
the type of position features are limited and relationships between regions cannot be specified.

The third way is recommendation.
The system automatically infers and recommends relevant spatial constraints 
from initial query and results.
The recommendation results are displayed in an understandable way 
(e.g., text, graphics), so that the user can intuitively select preferable recommendation results.
By using recommendation,
users can easily specify complex queries that describe relationships
between objects.
The idea is developed in Sec.~\ref{sec:recom}.


\subsection{Database Indexing}
\label{sec:index}
We first index the region-based CNN features of the image database 
by taking an approach similar to \cite{Hinami2016}, 
which consists of the following two steps.
{\bf (1) Extract features:}
Given a database that consists of $N$ images, 
for each image $I_i$ ($i=1,...,N$),
$n_i$ regions $R_{i,j}$ ($j=1,...,n_i$) are detected by performing a selective search~\cite{Uijlings2013}
($n_i\sim2000$).
A $D$-dimensional CNN feature $\mathbf{v}_{i,j}\in \mathbb{R}^D$ is then extracted
from each region $R_{i,j}$, as $\mathbf{v}_{i,j} = F_{rcnn}(I_{i}, \mathbf{r}_{i,j})$, 
where $\mathbf{r}_{i,j} = [x, y, w, h]$ is the bounding box of $R_{i,j}$, 
and $F_{rcnn}$ represents the trained region-based CNN feature extractor. 
We use the multitask Fast R-CNN feature,
as explained later in Sec.~\ref{sec:mt}. 
{\bf (2) Index features:}
All extracted features are indexed following IVFADC~\cite{Jegou2011}, which is based on an inverted index
and product quantization (PQ), a state-of-the-art indexing for 
the approximate nearest neighbor search 
(each feature is compressed into 128-bit codes by using PQ, 
and the number of codewords $k'=2^{14}$ is used as an inverted index).
Since IVFADC can approximately compute both the $L_2$ distance and inner product,
a nearest neighbor search and linear classification (i.e., maximum inner product search) 
can be handled in one system.

\subsection{Search Procedure}
Given $n_o$ objects as a query ($n_o=2$ in the example in Fig.~\ref{fig:pipeline}),
we first process the query of each object independently
and then merge the results into an image-level result.
The results are ranked by image-level score and presented to the user
after filtering them by the position constraints.
The details are as follows (each step corresponds to Fig.~\ref{fig:pipeline} (a)--(d)):

{\bf (a) Preprocess object-level queries:}
A category name or an image with a RoI is given as a query for each object.
When an image $I^q$ with a RoI $\mathbf{r}^q \in \mathbb{R}^4$ is given, 
we first extract feature $\mathbf{v}^q \in \mathbb{R}^D$ from the RoI: $\mathbf{v}^q=F_{rcnn}(I^q, \mathbf{r}^q)$.
When a name of category or attribute $t^q$ is given, 
we learn a linear SVM classifier by using the images of the category $t^q$
(e.g., a `person' classifier is learned using `person' images), where
the SVM weight vector $\mathbf{w}^q \in \mathbb{R}^D$ is used to score the regions.
We pre-compute a number of popular category classifiers offline 
in the same way as R-CNN~\cite{Girshick2014}
by using images in the COCO~\cite{Lin2014} and Visual Genome~\cite{Krishna2016}.
If the classifier of $t^q$ is has not been computed,
we train the SVM online by crawling images by using Google image search,
in a similar way to VISOR~\cite{Chatfield2012}.
The weights are cached so that learning does not have to be done again.

{\bf (b) Search regions:}
This part is also processed for each object $O_l$ independently ($l=1,...,n_o$).
Given the object-level query vectors obtained in the previous step, 
we retrieve the regions with high relevance scores for each object. 
When a feature vector $\mathbf{v}^q \in \mathbb{R}^D$ is given (when the query is an example image with an RoI), 
the relevance score of region $R_{i,j}$ is computed as 
$1 / \|\mathbf{v}^q - \mathbf{v}_{i,j}\|^2$.
When the weight of the classifier $\mathbf{w}_c^q \in \mathbb{R}^D$ is given (when the query is the name of object category), 
the relevance score of region $R_{i,j}$ is computed as 
$\langle \mathbf{w}_c^q, \mathbf{v}_{i,j} \rangle$.
In either case, we can immediately search the regions with high relevance scores by using IVFADC;
we select the inverted lists with the $k_s$ highest score ($k_s$=64) 
and compute the score between query and features in the selected lists by PQ.
When the attribute with the weight of classifier $\mathbf{w}^q_a \in \mathbb{R}^D$ is also given as a query, 
the relevance score $\langle \mathbf{w}^q_a, \mathbf{v}_{i,j}\rangle$ is also computed by 
using PQ and added to the region score.

{\bf (c) Merge object-level results:}
Image-level scores are computed by aggregating the object-level scores.
For each image $I_i$, the maximum region score is computed as an image score for each object ${O_l}$ 
and the image scores of all $n_o$ objects are summed into a final score of $I_i$ that is then used 
to rank the images.

{\bf (d) Filtering by position constraints:}
The value of each position feature (Fig.~\ref{fig:posfeat}) is computed for the search results,
and the search results are filtered using position constraints given as a query.
The constraints can be refined by interactive feedback,
which can be processed in real-time because only this stage (d)
is re-calculated on the client-side and involves no communication with the server.

\section{Spatial Relationship Recommendation} 
\label{sec:recom}
The position query of our system is represented as a set of position constraints.
Users can easily refine them because the constraints are unambiguously defined by equations.
However, their specification becomes more complicated as the number of constraints increases.
An alternative way of specifying a position query is by using text such as 
`A is at the top of the image' or `A is on the left of B'.
Although this way is intuitive, it is ambiguous and difficult to refine.
There is a trade-off between intuitiveness and unambiguousness, 
although both characteristics are important in query specification.

We hence developed a recommendation system that bridges the gap between 
{\it intuitive} and {\it unambiguous} query specifications.
Our system presents position query candidates with graphics or text
so that the user can intuitively choose one.
In addition, each recommended position query is represented as a set of position constraints
so that the user can easily refine their query.
This procedure enables users to intuitively specify and unambiguously refine the query.
We propose two types of recommendation (Fig.~\ref{fig:pipeline} lower right):
\begin{itemize}
\item{\bf Mining-based approach:} mining typical patterns of the object spatial positions
from the initial search results
\item{\bf Language-based approach:} predicting the spatial relationship between 
objects from a pair of object-level queries (text or image) on the basis of the language prior  
\end{itemize}

\begin{figure*}[t] 
\begin{center}
\includegraphics[width=1.00\linewidth]{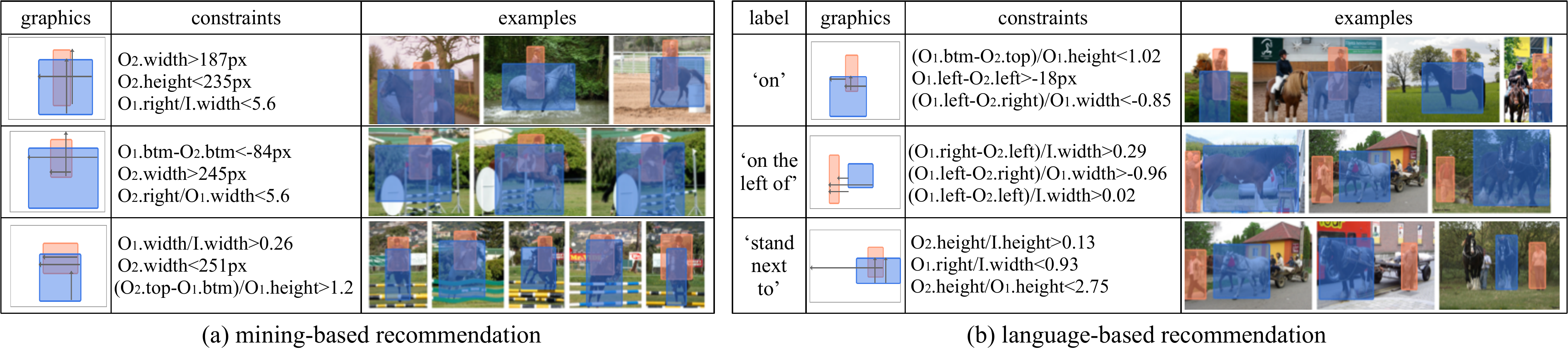}
\vspace{-5mm}
\caption{Examples of spatial relationship recommendation (query=`person', `horse').}
\vspace{-3mm}
\label{fig:recom_qual} 
\end{center} 
\end{figure*}

These two methods are complementary;
the mining-based approach depends on the database and not on the query, 
while the language-based approach depends on the query and not on the database.
Therefore, the mining-based approach can search for typical relationships that frequently occur
in the target database even if they cannot be represented in language 
(e.g., Fig.~\ref{fig:recom_qual}a shows three different contexts of `ride').
On the other hand, the language-based approach can recommend relationships that are rare in the database,
and the results are presented along with the label of relationship (e.g., `next to'), 
which allows users to make selections intuitively.

\begin{figure}[t] 
\begin{center}
\includegraphics[width=1.00\linewidth]{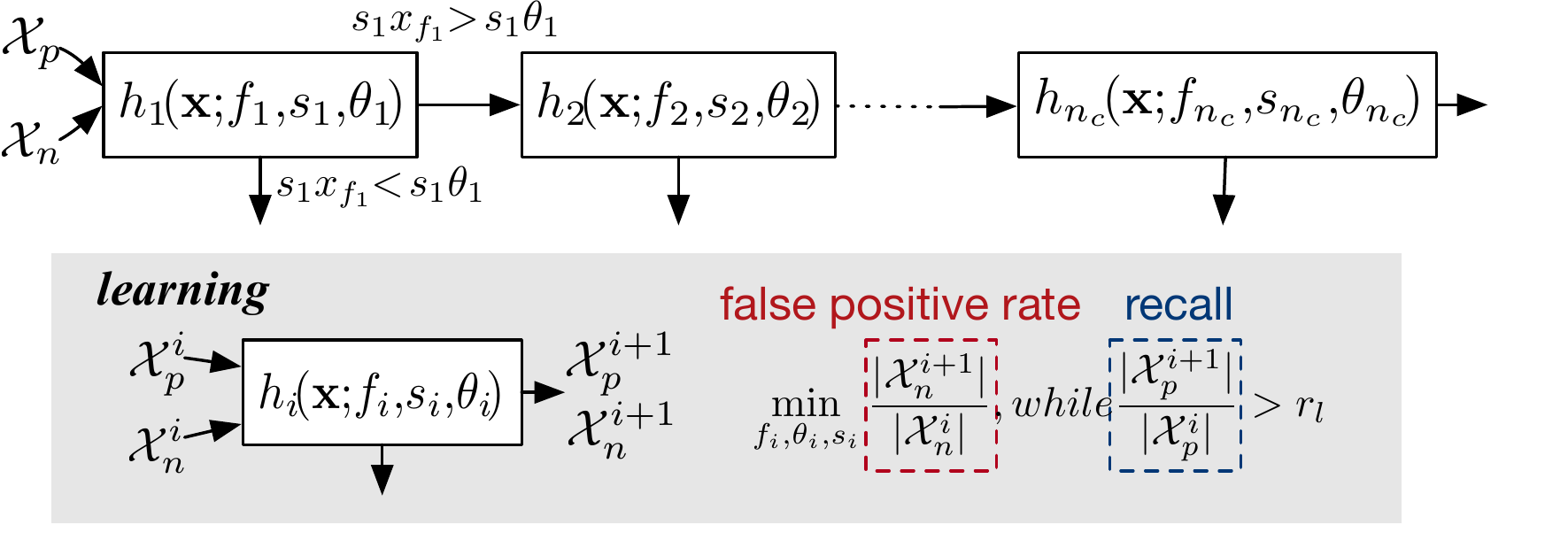}
\vspace{-5mm}
\caption{Illustration of position constraint learning.}
\vspace{-5mm}
\label{fig:cascade} 
\end{center} 
\end{figure}

\subsection{Mining-Based Recommendation}
\label{sec:recom1}
When users specify objects as a query, 
they want to search for images in which the objects are in a specific context.
For example, suppose we want to search for a ``person riding horse''.
Other results such as a ``person is standing next to horse'' become noise.
However, in existing search systems, including popular search engines,
it is difficult to make queries that focus on specific contexts of objects. 
Although some methods specify the relationship 
between objects by using language~\cite{Cao2015,Johnson2015a} (e.g., A is in front of B),
object contexts often cannot be represented in language.
To deal with this problem, we developed a way of automatically finding typical
spatial contexts of objects in database by mining the search results
that are used for making recommendations.
By mining search results, we can find object contexts that cannot be 
represented in language and are specific to the target database.

First, we find typical spatial patterns by clustering the initial search results 
on the basis of position features (described in Sec.~\ref{sec:method}).
The initial search results, each of them regarded as a point in a high-dimensional space 
defined by the set of all $n_p$ position features, are clustered by k-means (the number of cluster $K$=10),
where each cluster $\mathcal{C}_k$ ($k$=1,...,$K$) corresponds to one typical pattern (small clusters with $|\mathcal{C}_k| < 5$ are removed).
Second, because the clustering results do not always perfectly match the user's search intention,
we map them into position constraint queries so that user can interactively refine the suggested query.
The position constraints are learned for each cluster in order to distinguish samples in the target cluster from others (explained later).

Figure~\ref{fig:recom_qual}a shows the learned constraints and search results
with the query `person' and `horse'.
This approach retrieves images showing `person riding horse' in particular contexts.
A representative spatial layout of each cluster is also displayed (Fig~\ref{fig:recom_qual}a `graphics')
so that users can intuitively select recommendations; the sample closest to the centroids 
of the cluster is selected as a representative example.
The method to learn position constraints is as follows.

{\bf Learning of position constraints.}
The goal is to learn a set of position constraints that can be used to 
extract data in the target cluster $\mathcal{C}_k$
while filtering out the data in clusters other than $\mathcal{C}_k$. 
As shown in Fig.~\ref{fig:cascade}, the constraints can be regarded as a cascade classifier~\cite{Viola2001}, 
i.e., a degenerate decision tree,
where each constraint corresponds to a stage of the cascade.
Examples that satisfy all constraints are classified as positive in our case.
Each constraint function is denoted as $h_i(\mathbf{x}; f_i, \theta_i, s_i)$, 
which is parameterized by the type of position feature $f_i \in \{1,...,n_p\}$, 
threshold of feature $\theta_i$, and sign of inequality $s_i \in \{-1,1\}$;
the example that has position features $\mathbf{x}=[x_1,...,x_{n_p}]$ with $s_ix_{f_i}<s_i\theta_i$ is rejected by the constraint $h_i$.
The parameters of the constraints are learned in a way that is similar to cascade classifier learning.
In the training, the data in the target cluster are regarded as positive examples, and 
the data in other clusters are regarded as negative examples.
Formally, we learn a set of constraints $\mathcal{H}=\{h_1,...,h_{n_c}\}$ 
that separate positive examples $\mathcal{X}_p$ (=$\mathcal{C}_k$) from negative examples $\mathcal{X}_n = \{ \mathbf{x} | \mathbf{x} \in \bigcup_{l\neq k} \mathcal{C}_l\}$.
Each constraint is learned by minimizing the false positive rate while 
maintaining a recall of more than $r_l$ ($r_l = 0.96$ was used in this study).
The optimal type of feature $f_i$ and threshold $\theta$ is then found for each constraint.
$n_c$ constraints are learned in this way ($n_c=3$ is used unless otherwise specified).

\subsection{Language-Based Recommendation}
\label{sec:recom2}
The user intention can be estimated to some extent
from a query that contains a pair of object categories.
For example, given the query `person' and `bicycle', it is likely that the user wants to
search for images showing a person riding a bicycle.
Because some of such language-based relationships (e.g., ride, on the left of) 
are spatial, the results with a specific type of language-based relationship can be 
extracted from the search results by properly designing the spatial position constraints. 
Our language-based recommendation method estimates relationships on the basis of the language prior 
and provides corresponding position constraints,
where the correspondence between the relationships and constraints is learned
in advance on the dataset of visual relationship detection~\cite{Lu2016a}.

The language-based recommendation has two key components: 
1) predicting relationships from the query on the basis of the language prior 
and 2) learning position constraints for each relationship.
The relationship prediction is the same as used in Lu et al.~\cite{Lu2016a}.
Two input object category names are mapped to a word embedding space 
by using a pre-trained word2vec.
If an image is given as a query, its object category name is selected from
ImageNet 1000-class by predicting the class of the image with AlexNet~\cite{Krizhevsky2012}.
Two word2vec vectors are concatenated and input to the relationship classifier.
We used the language module in Lu et al.~\cite{Lu2016a} as the relationship classifier;
it learns a linear projection that maps concatenated word2vec 
vectors into likelihoods of relationships in the vocabulary $\mathcal{V}$ (e.g., `ride', `on', `next to').
Relationships with high likelihoods are selected as recommendations.

A set of position constraints corresponding to each relationship is learned using
the training set of the visual relationship detection (VRD) dataset~\cite{Lu2016a}.
This dataset includes annotations of object pairs and their relationships. 
An object category name and bounding box are annotated for each object.
To learn a set of position constraints of the relationship $r \in \mathcal{V}$, 
a set of positive features $\mathcal{X}_p$ is extracted from the object pairs with the relationship $v$ 
and a set of negative features $\mathcal{X}_n$ is extracted from relationships other than $v$.
Since the VRD dataset provides only one relationship for each object pair, 
the training data, especially the negative samples, may be very noisy.  
Note that the nominal number of object pairs may have multiple relevant relationships, 
while they are annotated only by one of such relationships;
hence, they can be potential noisy negative samples of the other relationships.
For example, an image of ``person rides a bicycle'', when annotated as `ride' but not as `on', 
this image is an potential noise of the relationship `on'.
In addition, the numbers of positives and negatives are unbalanced 
($|\mathcal{X}_p|\ll|\mathcal{X}_n|$).
These problems make it difficult to learn a standard classifier (e.g., SVM),
as is shown in the experiments in Sec.~\ref{sec:eval_recom}.
We learn the constraints from $\mathcal{X}_p$ and $\mathcal{X}_n$ by using 
the same algorithm as in Sec. \ref{sec:recom1}.
Because the algorithm guarantees the lower bound of the recall
(i.e., it guarantees the fraction of positive examples and 
minimizes the fraction of negative examples including nominal amount of noise),
the learned constraints can detect relationships with high recall despite these problems.
Examples of learned constraints and corresponding search results are shown in 
Fig.~\ref{fig:recom_qual}b.
Note that this language-based recommendation can be made to work for a larger number of objects 
by specifying multiple pairwise relationships
(e.g., $O_1$ ride $O_2$ and $O_2$ is next to $O_3$).


\section{Multitask CNN Feature} 
\label{sec:mt}
The purpose of this section is to learn a CNN feature that performs well in multiple tasks, 
which we call the multitask CNN feature, to use it in our system (${\bf v}_{i,j}$ in Sec.~\ref{sec:index}).
Our search system deals with multiple search tasks, namely,
instance search, object category search, and object attribute search,
although the optimal feature is different in each task.
Features that are effective in multiple visual search tasks have not received much attention 
because previous systems generally deal with different tasks independently.
However, a system that performs well at multiple tasks would be useful
in various situations, such as might a generic image search engine 
where users have very diverse search intentions.
While CNN features generalize well to multiple visual tasks,
their performance deteriorates if the tasks between the source and target are not similar,
such as between instance retrieval and category classification~\cite{Azizpour2015}.
Therefore, we investigated several multitask CNN architectures 
that can extract features that perform well in multiple tasks.

\begin{figure}[t] 
\begin{center}
\includegraphics[width=1.00\linewidth]{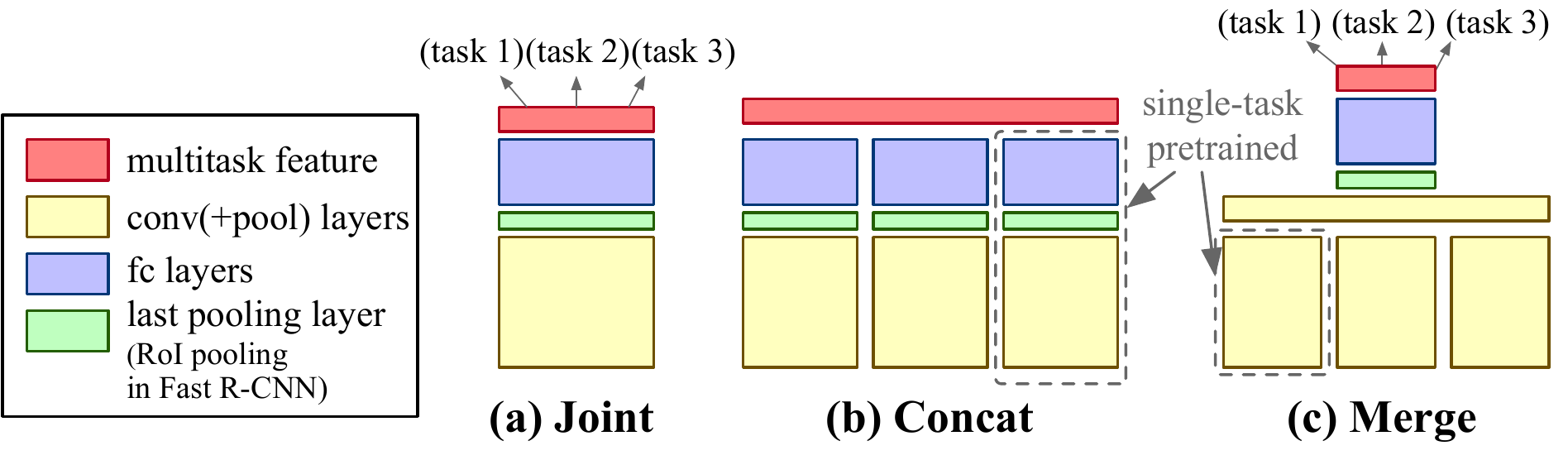}
\vspace{-7mm}
\caption{CNN architecture for the multitask feature.}
\vspace{-4mm}
\label{fig:mtcnn} 
\end{center} 
\end{figure}

{\bf Architectures.}
Figure~\ref{fig:mtcnn} shows three architectures of multitask CNN:
Joint, Concat, and Merge.
To obtain features that perform well in multiple tasks, 
we designed architectures that leverage the information in multiple tasks
by combining features trained on single-task and/or by jointly learning the multiple tasks.
Our architectures are based on AlexNet~\cite{Krizhevsky2012} or 
VGG~\cite{Simonyan2015},
which consists of a set of convolutional and pooling layers (Convs),
two fully connected layers (FCs), and an output layer for classification.
The architectures can be made compatible with the Fast R-CNN-based architecture
by simply replacing the last pooling layer (pool5) with a RoI pooling layer.
We removed the bounding box regression layer from the Fast R-CNN,
because we only wanted to extract features.

{\bf (a) Joint} is the same architecture as in the standard single-task model,
except for it having multiple classification layers,
where Convs and FCs are shared in all tasks.
Since Joint is only single-branch, its feature extraction 
is the fastest of all architectures. 
However, its training converges much more slowly than the training of other 
architectures because it learns all tasks jointly in a single branch.
{\bf (b) Concat} concatenates the features of each task's trained model.
This model learns each task separately in different branches;
i.e., it does not use multitask learning.
Its feature extraction is the slowest of the three architectures.
While it is the most accurate in classification because
it leverages the features of multiple tasks, 
its accuracy in nearest neighbor-based retrieval tasks is poor (see the experiments in Sec.~\ref{sec:eval_multi}).
{\bf (c) Merge} is a compromise between Joint and Merge, 
where separate Convs are used for each and FCs are shared among the tasks.
It merges conv5 feature maps of multiple tasks into one map by concatenating feature,
and reducing dimension with 1$\times$1 convolutional layer (we call it the merge layer) 
similar to the approach in \cite{Bell2016}.
This model learns Convs for each task separately and 
then learns the merge layer and FCs for all tasks jointly.
The multitask learning of Merge converges faster than Joint because 
it does not have to learn Convs.
Merge is more accurate than Joint because it uses separate Convs optimized for each task.

{\bf Training.}
Training the multitask feature consists of single-task and multitask training.
First, every branch of our models is pre-trained by using single-task training following 
the standard approach of every base network~\cite{Krizhevsky2012,Simonyan2015,Girshick2014,Girshick2015}.
If a pre-trained model is available, we can omit the single-task training.
Multitask training is then performed in the Joint and Merge models,
where the network is trained to predict multiple tasks.
In each iteration, a task is first randomly selected out of all tasks, 
and a mini-batch is sampled from the dataset of the selected task. 
The loss of each task is then applied to its classification layer and 
all Convs and FCs.
The Merge model pre-trains the merge layer alone by freezing other layers
and then fine-tunes the merge layer and FCs by freezing the Convs.

\section{Experiments}
\subsection{Spatial Relationship Recommendation} 
The experiments described in this subsection evaluate the position constraints 
used in spatial relationship recommendation.
As the goal of spatial relationship recommendation is to bridge the gap between intuitive specification and unambiguous refinement of spatial position queries, each recommended spatial relationship was represented by position constraints.
The experiments tested whether the learned position constraint could extract appropriate spatial relationships or not.

\label{sec:eval_recom}
{\bf Mining-based recommendation.}
We evaluated the accuracy of position constraint learning in mining-based recommendation.
We defined the initial clustering result obtained by k-means as the ground truth 
and determined whether position constraints could be learned to reproduce the clustering result.
As explained in ~\ref{sec:recom1},
position constraints were learned for each cluster as a cascade-style classifier by regarding 
the data in the cluster to be positive. 
Precision, recall, and F-value of the learned classifier were computed for each cluster, and 
the mean values over all clusters were used as performance measures of the position constraint learning.
Table~\ref{tab:clst_recom} lists these measures for different numbers of constraints $n_c$.
We used the PASCAL VOC dataset as the image database 
and used various combinations of two categories in PASCAL as the query.
The table shows mean values over all queries (190 category pairs in total).
It shows that the position constraints can accurately reproduce the clustering results
in high accuracy even with only three constraints (87\% in mean F-value).
The examples in the supplementary video demonstrate 
the effectiveness of this recommendation.

{\bf Language-based recommendation.}
We determined whether the position constraints corresponding to each language-based relationship (e.g., ride, on)
can accurately extract only the target relationship.
The constraints of relationships are learned with the VRD training set~\cite{Lu2016a},
which is evaluated with the VRD test set in a similar way to that of the mining-based approach. 
The recall and selectivity (\# of detected data / \# of all data) were used as the measures. 
Precision was not used because the VRD dataset sometimes has 
potential noisy negative samples (explained in Sec.~\ref{sec:recom2}), 
and detection of such samples is heavily penalized in the precision evaluation.
In addition, relationships with fewer than ten test samples were 
removed from the test set.
We tested a set of baselines: 
1) Visual relationship detection (VRD) in \cite{Lu2016a} was the main competitor.
VRD computes the relationship score when given two objects, 
as in the predicate detection task of \cite{Lu2016a}.
We used this score to obtain the detection results for each relationship, 
where the threshold of the score was determined from the training set to achieve 90\% recall.
2) We tested the combination of our position feature and standard classifiers
(linear SVM and random forest (RF)).
The classifiers were learned from the same positive and 
negative position features as our approach.
3) We tested another version using a linear SVM adapted to our task (SVM (adapt)); 
the threshold of the score was determined so as to achieve 90\% recall on the training data.

\begin{figure}[t] 
\begin{center}
\includegraphics[width=1.00\linewidth]{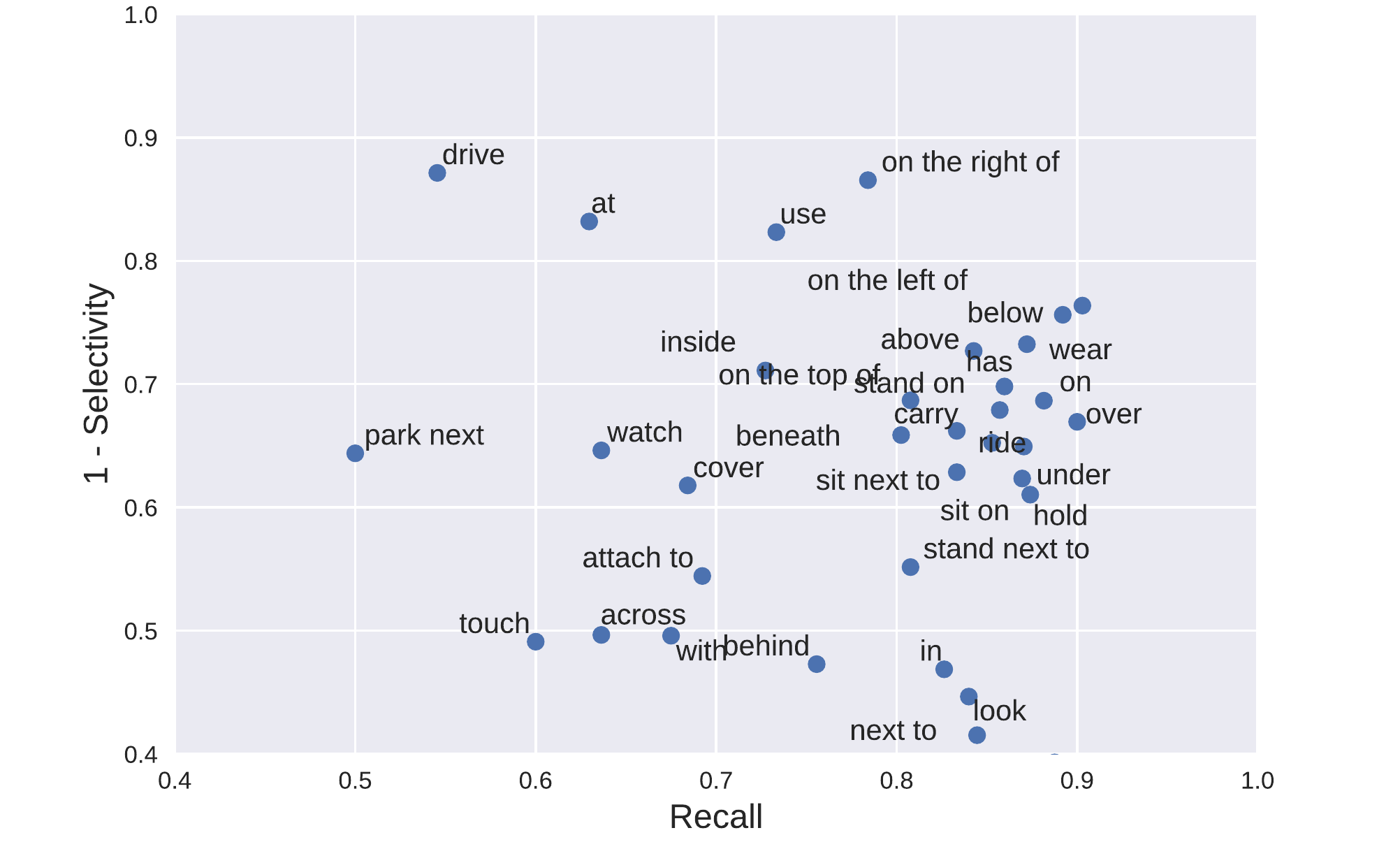}
\vspace{-6mm}
\caption{Results of language-based query recommendation.}
\label{fig:recom2_2} 
\vspace{-4mm}
\end{center} 
\end{figure}

Table~\ref{tab:recom2_1} shows the recall, selectivity, 
and harmonic mean of recall and 1 - selectivity 
of relationship detection.
The mean performance for all relationships is shown on the left 
(All relationships), and
the performance for only spatial relationships 
(e.g., on the left of, below) is shown on the right (Spatial only).
Although our method performed almost as well as VRD on ``All relationships'', 
it significantly outperformed the baseline in terms of spatial relations.
In addition, the combination of position features and the standard classifier 
did not achieve high recall.
Our approach learned a constraint that maintained high recall by 
determining a lower bound of recall that was effective in this problem.
In addition, our method based on position features is faster at 
both feature computation and classification than VRD is.
Figure~\ref{fig:recom2_2} plots the recall and selectivity of 
every relationship.
It shows that spatial relations such as `on the right of' and 
`below' are accurately detected by using the constraints,
while the relationships that are not much related to spatial position 
(e.g., `touch' and `look') cannot be detected by using the constraints.


\begin{table}  
\vspace{-2mm}
\caption{Results of clustering-based query recommendation.}
\vspace{-2mm}
\begin{center} 
\begin{tabular}{@{}l c c c c c@{}} \toprule
           &$n_c=1$&$n_c=2$&$n_c=3$&$n_c=4$&$n_c=5$\\
\midrule                                   
Mean precision  & 59.0  & 80.9  & 89.1  & 93.3  & 95.5 \\
Mean recall     & 93.0  & 90.6  & 86.5  & 82.7  & 79.4 \\
Mean F-value    & 69.6  & 84.7  & 87.2  & {\bf 87.3}  & 86.3 \\
\bottomrule
\end{tabular} 
\end{center} 
\label{tab:clst_recom} 
\end{table} 

\begin{table} 
\vspace{-2mm}
\caption{Relationship detection performance.} 
\vspace{-2mm}
\begin{center} 
\begin{tabular}{@{}l l c c c c c c@{}} \toprule
& & \multicolumn{3}{c}{{\bf All relationships}} & \multicolumn{3}{c}{{\bf Spatial only}}\\ 
\cmidrule(lr){3-5}\cmidrule(l){6-8}
Method            &                      & Rec. & Sel. & Har. & Rec. & Sel. & Har.   \\
\midrule                                     
Ours        & $n_c=2$                    & 84.3 & 47.0 & 65.0 & 88.9 & 64.9 & 60.9 \\
(posfeat.   & $n_c=3$                    & 78.2 & 38.5 & 68.8 & 84.5 & 72.4 & 76.3 \\
 +consts.)  & $n_c=4$                    & 73.7 & 32.9 & 70.3 & 80.5 & 76.3 & {\bf 76.8} \\
            & $n_c=5$                    & 68.0 & 29.0 & 69.5 & 75.4 & 23.4 & 76.0 \\
\midrule                                    
\multicolumn{2}{@{}l}{VRD~\cite{Lu2016a}}& 69.3 & 27.7 & {\bf 70.8} & 60.4 & 42.6 & 58.8 \\
\midrule
\multicolumn{2}{@{}l}{posfeat. + RF}     & 5.0  & 0.9  & 9.5  & 11.8 & 2.4  & 21.0 \\
\multicolumn{2}{@{}l}{posfeat. + SVM}    & 9.6  & 6.3  & 17.4 & 6.1  & 3.3  & 11.4 \\
\multicolumn{2}{@{}l}{posfeat. + SVM (adapt)} & 84.2 & 75.2 & 38.3 & 93.2 & 76.1 & 38.1 \\
\bottomrule
\end{tabular} 
\end{center} 
\label{tab:recom2_1} 
\end{table}
\subsection{Evaluation of Multi-task Feature}
\label{sec:eval_multi}
The experiments described in this subsection compare the performances of
several multitask and single-task CNN features on multiple tasks.
We first evaluate our method in a simple image classification and retrieval task 
using the {\it whole image feature}
(an evaluation using region-based features is described in the next subsection).
A multitask CNN was learned with three datasets, namely, 
ImageNet~\cite{Deng2009}, Places~\cite{Zhou2014}, 
and Landmarks~\cite{Babenko2014c}, 
which corresponded to the tasks of the object category, scene, 
and object instance classification.
We tested the performance of these tasks by using
PASCAL VOC 2007, CUB-200-2011~\cite{Gavves2013} (object classification), 
MIT indoor scenes~\cite{Quattoni2009}, SUN-397 scenes~\cite{Xiao2010} (scene classification), 
Oxford~\cite{Philbin2007}, and Paris~\cite{Philbin2008c} (instance retrieval).
In the classification task, we learned a linear SVM classifier on image features 
in the manner described in \cite{Azizpour2015}.
In the retrieval task, we simply performed a nearest neighbor search on the whole image feature space.
We post-processed the image feature with $l_2$-normalization, PCA-whitening, 
and $l_2$-normalization, following \cite{Jegou2012} for all tasks.
We used AlexNet as the base architecture, whose parameters were initialized 
using the models pre-trained on ImageNet.
All models are trained and tested with Chainer~\cite{tokui2015chainer}.

Table~\ref{tab:mt1} shows the performances of the features obtained by 
the multitask (Joint, Concat, and Merge) and single-task (ImageNet, Places, and Landmarks) models.
In Joint (freeze) and Merge (freeze), the parameters of Convs were frozen (not updated) 
in the multitask learning.
In Concat (4096), Concat features were reduced to 4096-dim by using PCA.
Following the standard metrics,
mean average precision (mAP) was used on the VOC, Oxford, and Paris datasets, and accuracy was used on 
the CUB, MIT, and SUN datasets.
We can see that Concat performs best in classification tasks 
(category and scene), while Joint and Merge performs best in
instance retrieval tasks.
These results imply that features from other unrelated tasks
become noise in the nearest neighbor search task,
whereas SVM classifiers can ignore unimportant features.
Merge performs better than Joint in the scene classification task,
which shows that even with multitask learning, Convs in a single 
branch has difficulty capturing features that are effective in multiple tasks.
The performance of Merge (freeze) is almost the same as that of Merge, 
which means fine-tuning Convs has no effect on the Merge model.

\begin{table}  
\vspace{-2mm}
\caption{Comparison of different CNN features.}
\vspace{-2mm}
\begin{center} 
\begin{tabular}{@{}l c c c c c c@{}} \toprule
& \multicolumn{2}{c}{{\bf object}} & \multicolumn{2}{c}{{\bf scene}}  & \multicolumn{2}{c}{{\bf instance}} \\ 
\cmidrule(lr){2-3}\cmidrule(lr){4-5}\cmidrule(l){6-7}
  & VOC & CUB & MIT & SUN & Oxford & Paris \\
\midrule
Joint           & 74.8 & 41.4 & 63.9 & 51.5 & 42.2 & 49.6 \\
Joint (freeze)  & 74.6 & 41.8 & 63.6 & 51.1 & 40.9 & 49.3 \\
Concat          & {\bf 76.2} & {\bf 43.3} & 70.6 & {\bf 57.3} & 36.5 & 43.3 \\
Concat (4096)   & 76.0 & 43.2 & {\bf 70.7} & 57.0 & 36.8 & 43.4 \\
Merge           & 76.0 & 41.5 & 67.1 & 53.1 & 42.8 & 48.1 \\
Merge (freeze)  & 76.0 & 41.5 & 67.2 & 53.1 & 42.8 & 48.1 \\
\midrule
ImageNet        & 74.5 & 42.2 & 53.1 & 40.2 & 30.8 & 35.2 \\
Places          & 72.7 & 22.6 & 67.1 & 53.6 & 30.8 & 39.5 \\
Landmarks       & 65.1 & 30.7 & 45.7 & 36.8 & {\bf 42.9} & {\bf 51.2} \\
\bottomrule
\end{tabular} 
\end{center} 
\label{tab:mt1} 
\vspace{-2mm}
\end{table} 
\begin{table}  
\vspace{-2mm}
\caption{Retrieval performance of our system. The same feature is used in the evaluation of every dataset.}
\vspace{-2mm}
\begin{center} 
\begin{tabular}{@{}l c c c@{}} \toprule
Method             &  VOC   & Oxford & Visual Phrase\\
\midrule                                   
Aytar et al.~\cite{Aytar2014}        & 27.5  & -    & -  \\
Large-scale R-CNN~\cite{Hinami2016}  & 50.0  & -    & -  \\
R-MAC~\cite{Tolias2015}              & -     & 66.9 & -  \\
Gordo et al.~\cite{Gordo2016}        & -     & {\bf 81.3} & -  \\
Visual phrase~\cite{Zhang2011}       & -     & -    & 38.0 \\
VRD~\cite{Lu2016a}                   & -     & -    & 59.2 \\
\midrule                                   
Ours (Joint)                         & 56.1 & 64.3 & 65.2 \\
Ours (Concat)                        & {\bf 57.7} & 57.6 & {\bf 66.3} \\
Ours (Merge)                         & 57.1 & 64.1 & 64.9 \\
\bottomrule
\end{tabular} 
\end{center} 
\label{tab:mt2} 
\end{table} 
\begin{figure*}[t] 
\begin{center}
\includegraphics[width=1.00\linewidth]{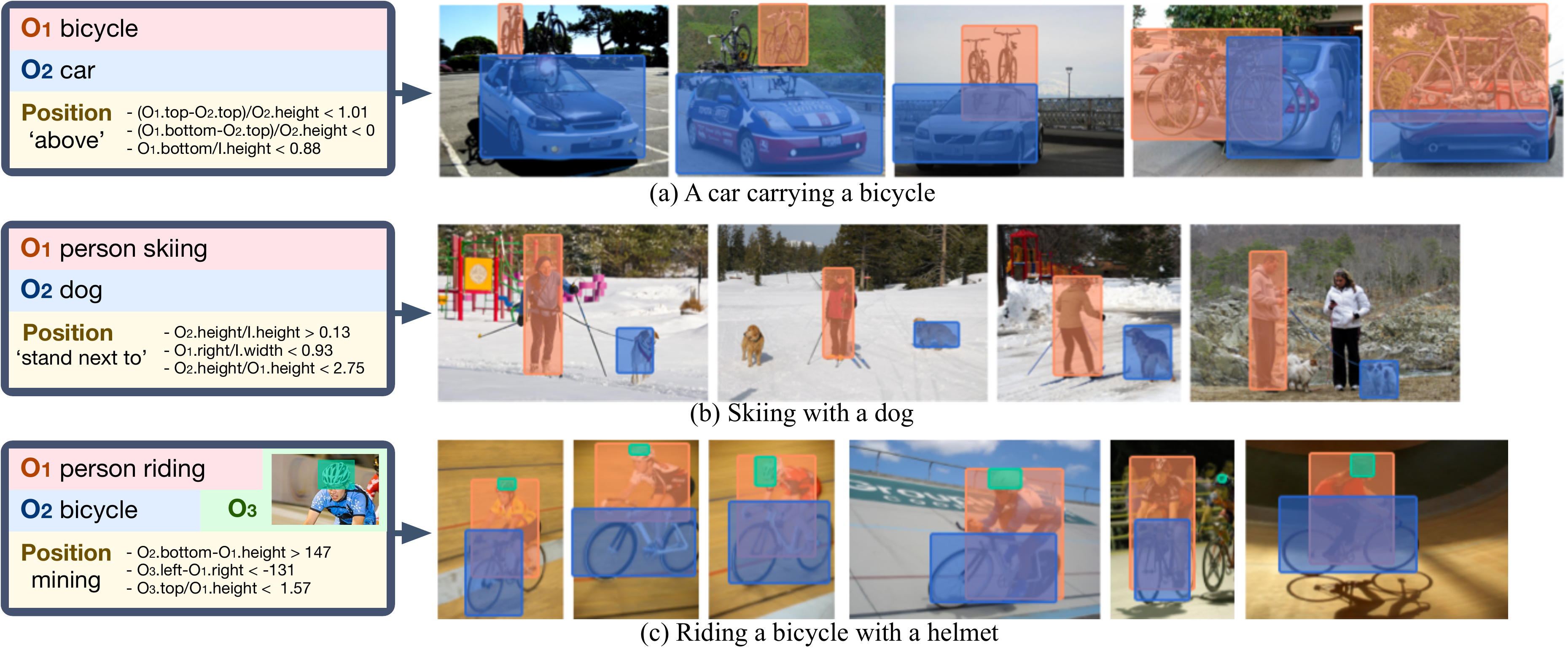}
\vspace{-6mm}
\caption{Examples of search results. Query consists of objects ($O_1$, $O_2$, and $O_3$) and 
their spatial constraints which are automatically suggested by recommendations.
Our system enable to capture various search intentions and retrieve accurately.}
\vspace{0mm}
\label{fig:example} 
\end{center} 
\end{figure*}

\subsection{Search Performance}
\label{sec:eval_frcn}


We evaluated the performance of our search system on 
datasets of three tasks: object category retrieval (PASCAL VOC), 
instance search (Oxford 5K~\cite{Philbin2007}), 
and visual phrase detection~\cite{Zhang2011}.
We trained the multitask Fast R-CNN model on the COCO, Landmarks, 
and Visual Genome datasets.
We used Visual Genome to learn attributes; 
72 frequently appearing attribute categories (color, action, etc.) were selected for training.
We used the VGG16 model as the base architecture.
We used the same hyper-parameters as in Fast R-CNN.
The object category retrieval task was evaluated with the same settings of \cite{Hinami2016}.
In the instance search, we extracted the query feature from the provided query image and RoI
and scored each image by using the distance between the query and its nearest feature of the image.
In the visual phrase detection, 
we evaluated the system on 12 phrases that represented a object1($o_1$)-relationship($r$)-object2($o_2$)
(e.g., person riding bicycle) as in ~\cite{Lu2016a}.
We manually converted each visual phrase into a query for our system; 
two objects were specified by category names ($o_1$ and $o_2$)
and position constraints is selected from the recommendation result without manual refinement.
Table~\ref{tab:mt2} compares the mAPs of our method with those of other state-of-the-art methods tuned for each task.
Although the other methods were tuned for each task, 
our method performed as well or better than them on all datasets.

Finally, let us discuss the search time of our system.
The query feature extraction and SVM training highly
depends on the CNN architecture and machine resources (e.g., CPU or GPU).
Although online SVM training is slow (around 5s per region) because of
image crawling and feature extraction, 
it only has to be done once per one category because we cache all trained classifiers.
The speed of region search part is the same as in \cite{Hinami2016};
130 ms on a 100K database for each object-level query.
The time taken by the recommendation part is negligible, 
less than 0.1 second.

Figure \ref{fig:example} shows examples of the results of our system,
which demonstrate that system can flexibly handle queries that have diverse intentions
and it produces promising results that match the query with high accuracy.
Position query are specified by recommendation.
More examples are provided in the supplemental video.


\section{Conclusion}
We presented region-based image retrieval (RBIR) system that supports 
semantic object specification and intuitive spatial relationship specification.
Our system supports multiple aspects of semantic object specification
by example images, categories, and attributes. 
In addition,
users can intuitively specify the spatial relationship between objects 
with system's recommendations and interactively refine the suggested queries. 
This semantic and spatial object specification 
allows users to make queries that match their intentions. 
The examples shown in this paper demonstrate the effectiveness of our system 
as well as the potential of RBIR.
RBIR provides a way for users to access the detailed properties of objects 
(e.g., category, attribute, and spatial position) in image retrieval and
we expect it to bring forth promising and exciting applications.

{\bf Acknowledgements:}
This work was supported by JST CREST JPMJCR1686
and JSPS KAKENHI 17J08378.

\bibliographystyle{ACM-Reference-Format}
\balance
\bibliography{RBIR} 

\end{document}